\documentstyle[12pt]{article}
\begin{document}

\newcommand{\ad}{{\rm ad}}
\newcommand{\Ad}{{\rm Ad}}
\newcommand{\ti}[1]{\tilde{#1}}

\newcommand{\al}{\alpha}
\newcommand{\th}{\theta}
\newcommand{\vth}{\vartheta}
\newcommand{\be}{\beta}
\newcommand{\lm}{\lambda}

\newcommand{\La}{\Lambda}
\newcommand{\D}{\Delta}
\newcommand{\de}{\delta}
\newcommand{\ka}{\kappa}
\newcommand{\ve}{\varepsilon}
\newcommand{\vf}{\varphi}
\newcommand{\G}{\Gamma}
\newcommand{\ip}{\hat{\upsilon}}
\newcommand{\Ip}{\hat{\Upsilon}}
\newcommand{\ga}{\gamma}
\newcommand{\li}{\lim_{n\rightarrow \infty}}
\newcommand{\mat}[4]{\left(\begin{array}{cc}
{#1}&{#2}\\{#3}&{#4}\end{array}\right)}
\newcommand{\si}[1]{\sigma _{#1}}
\newcommand{\beq}[1]{\begin{equation}\label{#1}}
\newcommand{\eq}{\end{equation}}
\newcommand{\beqn}[1]{\begin{eqnarray}\label{#1}}
\newcommand{\eqn}{\end{eqnarray}}
\newcommand{\AJM}{Amer.  J. Math.}
\newcommand{\AM}{Adv. Math.}
\newcommand{\p}[1]{\partial} 
\newcommand{\TAMS}{Trans.  Amer.  Math. Soc. }
\newcommand{\TMMS}{Trans.  Mosc. Math. Soc. }
\newcommand{\IM}{Inv. Math. }
\newcommand{\PL}{ Phys. Lett }
\newcommand{\NP}{Nucl. Phys.}
\newcommand{\JMP}{ J. Math. Phys. }
\newcommand{\JDG}{ J.\ Diff.\ Geom. }
\newcommand{\ANM}{ Ann.  Math.}
\newcommand{\DAN}{Dokl. Akad. Nauk. }
\newcommand{\IAN}{Izv.Akad.Nauk.SSSR  }
\newcommand{\UMN }{Usp. Math.Nauk }
\newcommand{\CMP}{Commun.\ Math.\ Phys. }
\newcommand{\TMF}{Theor.\ Math.\ Phys. }
\newcommand{\CQG}[1]{Class.\ Quant.\ Grav. }
\newcommand{\FAA}{Funct. Analys. Appl. }
\newcommand{\JP}[1]{J.\ Phys. }
\newcommand{\JA }[1]{ J. Algebra }
\newcommand{\JFA} {J. Funct. Anal. }
\newcommand{\MPL}[1] { Mod. Phys. Lett. }
\newcommand{\IJMP}[1] { Int. J. Mod. Phys. }
\newcommand{\FP}[1] {Fortschr.der Phys. }
\newcommand{\RMAP}[1] { Rep.\ Math.\ Phys. }
\newcommand{\PRL}[1]{Phys.\ Rev.\ Lett. }
\newcommand{\AP}[1]{Ann.\ Phys. }
\newcommand{\PTP}[1]{ Prog.\ Theor.\ Phys. }
\newcommand{\SPTP}[1]{Suppl.\ Prog.\ Theor.\ Phys. }
\newcommand{\PR}[1]{Phys.\ Rev. }
\newcommand{\PREP }[1]{Phys.\ Reports }
\newcommand{\NC}[1]{Nuovo \ Cim. }
\newcommand{\NCL }[1]{Nuovo\ Cim.\ Lett. }

\begin{flushright}
 ITEP-MO-93/3 \\
INFN FE-11-93 \\
\end{flushright}
\vspace{10mm}
\begin{center}
{\Large Liouville Quantum Mechanics on a Lattice}\\
 {\Large from Geometry of Quantum Lorentz Group}\\
\vspace{5mm}
 M.A.Olshanetsky \footnote{e-mail OLSHANEZ@VXDESY.DESY.DE} \\
{\em ITEP,117259 Moscow,Russia}\\
and \\
V.-B.K.Rogov\\{\em MIIT, 103055, Moscow, Russia} \\
\vspace{5mm}
September 1993\\
\end{center}
\begin{abstract}
We consider  the quantum Lobachevsky space ${\bf L}_q^3$, which
 is defined as subalgebra  of the Hopf algebra
 ${\cal A}_q(SL_2({\bf C}))$. The Iwasawa decomposition
of ${\cal A}_q(SL_2({\bf C}))$ introduced by Podles and Woronowicz
 allows to consider the quantum
 analog of the horospheric coordinates on ${\bf L}_q^3$.
 The action
of the Casimir element, which belongs to the dual to  ${\cal A}_q$
quantum group  $U_q(SL_2({\bf C}))$, on some subspace in
 ${\bf L}_q^3$ in these coordinates leads to a second order difference
operator on the infinite one-dimensional lattice. In the continuos limit
 $q\rightarrow 1$
it is transformed into
the Schr\"{o}dinger Hamiltonian, which describes zero modes  into the Liouville
field theory (the Liouville quantum mechanics).
We calculate the spectrum (Brillouin zones) and the eigenfunctions
 of this operator. They are
$q$-continuos Hermit polynomials, which are particular case of the Macdonald
or Rogers-Askey-Ismail polynomials. The scattering in this problem corresponds
to the scattering of first two level dressed excitations in the $Z_N$ Baxter
 model in the
very peculiar limit when the anisotropy parameter $\ga$ and
 $N~\rightarrow\infty$, or, equivalently, $(\ga, N)\rightarrow 0$.
\end{abstract}

	\section{Introduction}

 There are a lot of interrelations between integrable one-dimensional
 systems of particles and integrable 2d field theories. One of them is
 the similarity in dynamics between solitons and the  classical
 Calogero-Moser-Sutherland-Toda particles and their r
generalizations \cite{R1}.
On a quantum level the similar phenomena has been observed recently
 in \cite {Z,FZ}
\footnote{Quantum dynamics was discussed partly in \cite{R1}}.
In particular, it was
discovered there that the scattering of some special excitations in
 the $Z_2$-Baxter
model coincides with the scattering, which is defined by asymptotics of the
Macdonald polynomials for the root system $A_1$ \cite{M}.
They depend on two parameters and  in the simplest case define the  usual
 zonal spherical functions on
reducible symmetric spaces of rank one.

Recall that the  zonal spherical functions are defined as a normalized
 eigenfunctions of the Laplace-Beltrami operator on a symmetric space,
which are invariant with respect to a stationer subgroup and
 have a free asymptotic.
 In other words, the Laplace-
Beltrami operator in the spherical coordinates coincides up
 to a simple gauge transformation
 with the generalized Calogero-Moser-Sutherland potential \cite{OP}.
The  corresponding Jost
function is the famous Harish-Chandra $c$-function \cite{HC},
 which is factorizable
\cite{GK}. This fact is an agreement with the complete integrability
 of the Calogero-
Mozer-Sutherland systems .
In particular, the Jost functions  for a symmetric noncompact space of rank one
  give rise to the scattering
of a quantum particle on a semi-line on the Calogero-Moser potential
 $\frac{g^2}{\sinh ^2x}$ .
The same $S$-matrix arises in the scattering of two kinks in the XXX-model.
In the similar fashion XXZ-model corresponds to the quantum symmetric space
of rank one. But in this case the continuous
 semi-line is replaced on a semi-indefinite
lattice.

The open quantum Toda Hamiltonian can be derived in
 the similar way \cite{OP,STS}.
To this end the same Laplace-Beltrami operator is considered
 in the horospherical coordinates.
  If its eigenfunctions are independent
on the horospherical "angular" coordinates then the
 Laplace-Beltrami operator is reduced to a second order differential
 operator with constant coefficients.
 But if the functions have nontrivial multiplier
when their arguments are shifted along horospheres
 ( more exactly, they belong to
a representation, induced by a character of a nilpotent subgroup),
 then the Laplace-Beltrami operator
acquires a nontrivial potential term,
 which is nothing else as the open Toda potential.
The corresponding eigenfunctions are called the Whitteker
 functions \cite{J,Sc}.
It was proved in \cite{STS} that the corresponding
 Jost functions are also factorizable.
The simplest $SL_2$ case describes the scattering on
 the Liouville potential $e^{-2x}$.

We can consider this model as a simplified version
 of the  ubiquitous Liouville
field theory,  for $\phi (\sigma,t)$
 (the zero mode $x(t)=
\phi (\sigma,t)|_{\sigma=0} $)
, which describes 2D induced gravity in the conformal gauge.
Then, $e^{-2x}$ defines the circumference of 1d "universe".
 The quantum mechanics of the zero modes sheds light on
 the spectrum of the full theory in
the quasiclassical approach \cite{GM}.

Our aim is to put this theory on a lattice. In other words,
 $x$ will take discrete values.
There are a few reasons to do it. First of all, introducing
 the new parameter - the step
of a lattice, is equivalent to the generalizations of corresponding
 spin chain Lagrangians.
Next, this model in the 2d version modifies the 2d gravity
 in the similar fashion as a
finite group approximation of gauge groups in the Yang-Mills theories.
 This modification
of 2d  gravity can be in principle useful in the quantization procedure.
Finally, classical solutions   of  models
 on lattices may lead presumably to deformed tau
 functions connected with new integrable hierarchies, as well as to
partition functions of some topological theories.

 Here we consider the very simple quantum theory, which
turns out to be exactly solvable.
To derive the model we proceed to "the second quantization" of the Liouville
 quantum mechanics using the formalism of quantum groups.
Namely, we consider the quantum Lobachevsky space ${\bf L}_q^3$.
It is defined as a subalgebra of the Hopf algebra
 ${\cal A}_q\supset{\bf L}_q^3$,
 which is dual to the quantum Lorentz group
$U_q(SL_2({\bf C}))$ .  The algebra ${\bf L}_q^3$ is equipped with a
right $U_q(SL_2({\bf C}))$-module structure.
 In ${\bf L}_q^3$  exists an analog of the
 horospherical coordinates,  which are connected
 with the Iwasawa decomposition of ${\cal A}_q$ \cite{PW}.
 We calculate the Casimir element
$\Omega_q\in U_q(SL_2({\bf C})$ in these  coordinates. It is possible to
reduce the operator to  subspace, which is similar to the space of the induced
representation of the nilpotent subgroup in the
 classical situation. The quantum horospherical variables are
separated in this subspace. The operator $\Omega_q$ becomes classical second
order difference operator on a one-dimensional
 lattice with the Liouville "wall".
 It differs slightly from relativistic Liouville,
 introduced in \cite{R}.  In the  limit,
when a step of the lattice vanishes
, $\Omega_q|_{(q\rightarrow 1)}$ coincides with the Liouville Hamiltonian.
 The quantum Whitteker functions are the $q$-continuous Hermit polynomials,
which are particular case of the Rogers-Askey-Ismail polynomials \cite{GR},
or the Macdonald polynomials for the $A_1$ root system \cite{M}. Thus,
the Macdonald polynomials serve also to define the Whitteker functions,
 as well as
the zonal spherical functions.

The
Harish-Chandra c-function gives rise to the $S$-matrix, which coincides with
the $S$- matrix for scattering of first two levels in the $Z_N$-Baxter model
\cite{B,G} in the limit $N,\ga\rightarrow\infty$, where $\gamma$
is an anisotropy parameter, or, equivalently, $\gamma,~N\rightarrow 0$.

\section{Classical Case}
Let ${\bf L}^3=SU_2\backslash SL_2({\bf C})$ be
 is a homogeneous space of the second
order unimodular Hermitian positive definite
 matrices, which is a model of the classical
Lobachevsky space.
Let
$$g=\mat{\al}{\be}{\ga}{\de},~\al\de-\be\ga=1.$$
Then any $x\in{\bf L}^3$ can be represented as
\beq{1.2}
x=g^{\dagger}g=\mat{\bar{\al}\al+\bar{\ga}\ga}{\bar{\al}\be+\bar{\ga}\de}
{\bar{\be}\al+\bar{\de}\ga}{ \bar{\be}\be+\bar{\de}\de}.
\eq

The Iwasawa decomposition
\beq{1.1}
g=kb, ~g\in  SL_2({\bf C}),~ k\in SU_2,~b\in AN-{\rm Borel~subgroup}
\eq
allows to define the horospherical coordinates on ${\bf L}^3$.
If
$$b=\mat{h}{hz}{0}{h^{-1}},$$
 then from (\ref{1.2})
\beq{1.3}
x=b^{\dagger}b=\mat{\bar{h}h}{\bar{h}hz}{\bar{z}\bar{h}h}{
 \bar{z}\bar{h}hz+(\bar{h}h)^{-1}}.
\eq
The triple $(H=\bar{h}h,z,\bar{z})$ is uniquely determined by $x$. It is called
the horospherical coordinates of  $x$.
 It follows from (\ref{1.2}) and (\ref{1.3}) that
$$H=\bar{\al}\al+\bar{\ga}\ga,$$
$$Hz=\bar{\al}\be+\bar{\ga}\de,$$
$$\bar{z}H=\bar{\be}\al+\bar{\de}\ga.$$

Let for
$$g=\mat{A}{B}{C}{D}\in SL_2({\bf C})$$
$d_A,d_B,d_C$ and $d_D$ be the Lie operators of the right shift on ${\bf L}^3$.
In the horospherical coordinates they take the form
$$d_A=\frac{1}{2}H\partial_H-z\partial_z,~~d_D=-d_A,$$
\beq{1.4}
d_B=\partial_z,
\eq
$$d_C=Hz\partial_H-z^2\partial_z+H^2\partial_{\bar{z}}.$$
The second Casimir
\beq{1.6}
\Omega=d_A^2+d^2_D+d_Bd_C+d_Cd_B
\eq
in the horospherical coordinates takes the form
\beq{1.7}
\Omega=\frac{1}{2}H^2\partial^2_H+\frac{3}{2}
 H\partial_H+2H^{-2}\partial^2_{z\bar{z}}.
\eq
Consider the eigenvalue problem
\beq{1.8}
\frac{1}{2}\Omega~
 \Phi_{\lm}(H,z,\bar{z})=-(\lm^2+\frac{1}{4})\Phi_{\lm}(H,z,\bar{z}),
\eq
and put
$$\Phi_{\lm}(H,z,\bar{z})=H^{-1}\exp i\mu(z+\bar{z})\phi_{\lm}(H),$$
 where
  $\exp i\mu(z+\bar{z})$ can be consider as a unitary character of the
nilpotent
 subgroup $N$.
 Then for $x=\frac{1}{2}\log H$ and  $\Psi_{\lm}(x)=\phi_{\lm}(H)$
 (\ref{1.8}) is transformed to
\beq{1.10}
(-\frac{1}{4}d^2_{xx}+4\mu^2e^{-4x})\Psi_{\lm}(x)=4\lm^2\Psi_{\lm}(x).
\eq
The solution to this equation
\beq{1.11}
\Psi_{\lm}(x)=K_{2i\lm}(2\mu e^{-2x})
\eq
is the Bessel-Macdonald function with the asymptotic behavior
\beq{1.12}
K_{2i\lm}(2\mu e^{-2x})\sim\frac{\pi}{\sin 2\pi \lm}
(\frac{(\mu e^{2x})^{-2i\lm}}{\G(1-2i\lm)}
+\frac{(\mu e^{2x})^{2i\lm}}{\G(1+2i\lm)}).
\eq
This solution is the so called Whitteker function for $SL_2({\bf C})$
\cite{J,Sc}.
The two-body $S$-matrix
\beq{1.13}
S(\lm)=\frac{\G(1+2i\lm)}{\G(1-2i\lm)},
\eq
which is obtained from (\ref{1.12}), desribes the scattering of a quantum
particle on the Liouville "wall" $e^{-4x}$.

\section{Quantum Lobachevsky Space}
Let ${\cal A}_q(SL_2({\bf C}))
 ~(0<q\leq 1)$ be the algebra of functions on  $SL_2({\bf C})$
\cite{PW}, which
is defined as the factor algebra
 of the free associative $\bf{C}$-algebra with generators
$\al,\be,\ga,\de$ , with an antiinvolution $*:{\cal A}_q \rightarrow{\cal A}_q,
{}~(ab)^*=b^*a^*$ and the following
 relations
$$
\al\be=q\be\al,~\al\ga=\ga\al,~\be\de=q\de\be,~\ga\de=q\de\ga,
{}~\be\ga=\ga\be,
$$
$$\al\de-q\be\ga=1,~\de\al-q^{-1}\be\ga=1,$$
$$\be\al^*=q^{-1}\al^*\be+q^{-1}(1-q^2)\ga^*\de,$$
$$\ga\al^*=q\al^*\ga,~\de\al^*=\al^*\de,~\ga\be^*=\be^*\ga,$$
\beq{2.1}
\de\be^*=q\be^*\de-(1-q^2)\al^*\ga,
\eq
$$\de\ga^*=q^{-1}\ga^*\de,~\al\al^*=\al^*\al+(1-q^2)\ga^*\ga,$$
$$\be\be^*=\be^*\be+(1-q^2)(\de^*\de-\al^*\al)-(1-q^2)^2\ga^*\ga,$$
$$\ga^*\ga=\ga\ga^*,~\de\de^*=\de^*\de-(1-q^2)\ga^*\ga.$$
The rest commutation relations can be read off from the rule $(ab)^*=b^*a^*.$

We cast the generators into the matrix form
$$w=\mat{\al}{\be}{\ga}{\de}, ~w^*=\mat{\al^*}{\be^*}{\ga^*}{\de^*}.$$

With the comultiplication $\D:
{\cal A}_q \rightarrow{\cal A}_q\otimes {\cal A}_q$
$$\D\mat{\al}{\be}{\ga}{\de}=\mat{\al}{\be}{\ga}{\de}\otimes
\mat{\al}{\be}{\ga}{\de},$$
the antipode $S:{\cal A}_q \rightarrow{\cal A}_q$
$$S\mat{\al}{\be}{\ga}{\de}=\mat{\de}{-q^{-1}\be}{-q\ga}{\al},$$
and the counit $\ve:{\cal A}_q \rightarrow{\bf C}$
$$\ve\mat{\al}{\be}{\ga}{\de}=\mat{1}{0}{0}{1}$$
${\cal A}_q $ becomes a Hopf algebra.
In fact, it is a *-Hopf algebra, since
$$(\D(a))^*=\D(a^*),$$
and
\beq{2.2}
S\circ *\circ S\circ *={\rm id}.
\eq

We define the $*$-Hopf subalgebra ${\cal A}_q(SU_2)$ by the generators
\beq{2.3}
{\cal A}_q(SU_2)=\{w_c=\mat{\al_c}{-q\ga^*_c}{\ga_c}{\al_c^*}\}
\eq
and the relations
$$\al^*_c\al_c+\ga^*_c\ga_c=1,~~\al_c\al_c^*+q^2\ga^*_c\ga_c=1,$$
$$\ga_c^*\ga_c=\ga_c\ga^*_c,~~\al_c\ga_c^*=q\ga_c^*\al_c,$$
$$\al_c\ga_c=q\ga_c\al_c.$$
Then
\beq{2.3a}
w^*_cw_c=\mat{1}{0}{0}{1}.
\eq
In the similar way
\beq{2.4}
{\cal A}_q(AN_q)=\{w_d=\mat{h}{z}{0}{h^{-1}}\},
\eq
$$hh^*=h^*h,~hz=qzh,~zh^*=q^{-1}h^*z,$$
$$zz^*=z^*z+(1-q^2)((h^*h)^{-1}-h^*h).$$
The Iwasawa decomposition in the quantum context takes the form \cite{PW}
\beq{2.5}
w=w_cw_d,~w\in{\cal A}_q(SL_2({
\bf C})),~w_c\in{\cal A}_q(SU_2),~
w_d\in{\cal A}_q(AN_q).
\eq

Natural desription of  commutation relations (\ref{2.1}) can be obtained from
the construction of the quantum double . It was implemented
in \cite{JS}, where ${\cal A}_q(SL_2({\bf C}))$ is described   as
a special quantum double of ${\cal A}_q(SU_2)$, and  (\ref{2.2}) are derived by
means of corresponding $R$-matrix .

{\it The quantum Lobachevsky space} ${\bf L}_q^3$ is *-subalgebra of
${\cal A}_q(SL_2({\bf C}))$ generated by the bilinear constituents
\beq{2.6}
w^*w=\mat{\al^*\al+\ga^*\ga}{\al^*\be+\ga^*\de}{\be^*\al+\de^*\ga}
{\be^*\be+\de^*\de}=\mat{p}{s}{s^*}{r}.
\eq
Evidently, * acts as
$$p^*=p,~(s)^*=s^*,~r^*=r.$$
We don't need the explicit
 form of the commutation relations between $p,s,s^*$ and $r$ - they can be
derived from (\ref{2.1}).

Due to (\ref{2.3a}), (\ref{2.4}) and (\ref{2.6})
\beq{2.7}
p=H=h^*h=hh^*,~s=Hz,~s^*=z^*H,~r=z^*Hz+H^{-1}.
\eq
 The triple $(H,z,z^*)$ generates  the
horospherical coordinates in the algebra ${\bf L}_q^3$.
It folows from (\ref{2.1}) that
$$Hz=q^2zH,~H^{-1}z=q^{-2}zH^{-1},~z^*H=q^{-2}Hz^*,$$
\beq{2.8}
z^*H^{-1}=q^{-2}H^{-1}z^*,
\eq
$$zz^*=q^2z^*z+(q^2-1)(1-H^{-2}).$$

Consider now the complex associative algebra $U_q(SL_2(\bf{C}))$
with unit $1$,  generators
$A,B,C,D$ and the relations
$$AD=DA=1,~AB=qBA,~BD=qDB,$$
\beq{2.9}
AC=q^{-1}CA,~CD=q^{-1}DC,
\eq
$$[B,C]=\frac{1}{q-q^{-1}}(A^2-D^2).$$
In fact, it is the Hopf algebra, where
$$\D(A)=A\otimes A,~\D(D)=D\otimes D,$$
\beq{2.10}
\D(B)=A\otimes B+B\otimes D,
\eq
$$\D(C)=A\otimes C+C\otimes D,$$
$$\ve\mat{A}{B}{C}{D}=\mat{1}{0}{0}{1},$$
\beq{2.11}
S\mat{A}{B}{C}{D}=\mat{D}{-q^{-1}B}{-qC}{A}.
\eq

   There exists a nondegerate bilinear form $<u,a>: ~U_q\times {\cal A}_q
\rightarrow {\bf C}$ such that
$$<\D(u),a\otimes b>=<u,ab>,~<u\otimes v,\D(a)>=<uv,a>,$$
$$<1_U,a>=\ve_{\cal A}(a),~<u,1_{\cal A}>=\ve_{U}(u),~
<S(u),a>=<u,S(a)>.$$

 It takes the form on the generators
\beq{2.13}
<A,\mat{\al}{\be}{\ga}{\de}>=\mat{q^{\frac{1}{2}}}{0}{0}{q^{-\frac{1}{2}}},~
<D,\mat{\al}{\be}{\ga}{\de}>=\mat{q^{-\frac{1}{2}}}{0}{0}{q^{\frac{1}{2}}},
\eq
$$<B,\mat{\al}{\be}{\ga}{\de}>=\mat{0}{1}{0}{0},~
<C,\mat{\al}{\be}{\ga}{\de}>=\mat{0}{0}{1}{0}.$$
Moreover  $U_q(SL_2(\bf{C}))$ is *-Hopf algebra in duality, where the
involution
is defined by the pairing
\beq{2.12}
<u^*,a>=\overline{<u,(S(a))^*>}.
\eq

The element
\beq{2.14}
\Omega_q:=\frac{q^{-1}A^2+q^2D^2-2}{(q^{-1}-q)^2}+BC
\eq
is a Casimir element, since it commutes with
any $u\in U_q(SL_2(\bf {C}))$.

The right action of $u\in U_q(SL_2(\bf {C}))$ on ${\cal A}_q $ is defined as
\cite{Ko}
\beq{2.15}
a.u:=(u\otimes {\rm id})(\D(a)).
\eq
It is the algebra action:
\beq{2.15a}
a.(uv)=(a.u).v~,
\eq
which satisfies the "Leibnitz rule"
\beq{2.15b}
(ab).u=\sum_j(a.u_j^1)(b.u_j^2),~{\rm where}~\D(u)=\sum_ju_j^1\otimes u_j^2.
\eq
The left action is defined in the similar way.

The right action on the generators takes the form
$$\mat{\al}{\be}{\ga}{\de}.A=\mat{q^{\frac{1}{2}}\al}{q^{-\frac{1}{2}}\be}
{q^{\frac{1}{2}}\ga}{q^{-\frac{1}{2}}\de},
$$
\beq{2.16}
\mat{\al}{\be}{\ga}{\de}.B=\mat{0}{\al}{0}{\ga},
\eq
$$\mat{\al}{\be}{\ga}{\de}.B=\mat{0}{\al}{0}{\ga},$$
$$\mat{\al}{\be}{\ga}{\de}.D=\mat{q^{-\frac{1}{2}}\al}{q^{\frac{1}{2}}\be}
{q^{-\frac{1}{2}}\ga}{q^{\frac{1}{2}}\de}.$$

We will define now the right action of $U_q(SL_2(\bf{C}))$ on ${\bf L}_q^3$,
which endows the later with the structure of the right *-module.
For any $a\in{\bf L}_q^3$ define the normal ordering
using (\ref{2.1})
\beq{2.17}
:a:~:=\sum_k c_ka^*_{1,k}a_{2,k},
\eq
where $a^*_{1,k},~(a_{2,k})$ are monoms depending on $\al^*,\be^*,
\ga^*,\de^*~~(\al,\be,\ga,\de).$
Then the right action on ${\bf L}_q^3$, which will be denoted as
$(a).u$, is defined as follows
\beq{2.17a}
(a).u=\sum_k c_ka^*_{1,k}(a_{2,k}.u).
\eq
In particular, from (\ref{2.16})
$$(p,s,s^*,r).A=(p,0,s^*,0),~~(p,s,s^*,r).B=(o,p,0,s^*),$$
$$(p,s,s^*,r).C=(s,0,r,0),~~(p,s,s^*,r).D=(0,s,0,r).$$
To obtaine the action of generators of $U_q(SL_2(\bf {C}))$ on
 the horospheric
coordinates it is necessary to express them through  $\al^*,\be^*,
\ga^*,\de^*$ and $\al,\be,\ga,\de$, ordering them in correspondence with
(\ref{2.17}), and then to apply (\ref{2.17a}).
Two expressions have the normal order from the very beginning
 (see (\ref{2.6})
 and (\ref{2.7}))
\beq{2.18}
H=\al*\al+\ga^*\ga,
\eq
$$Hz=\al^*\be+\ga^*\de.$$
For $H^{-1}$ we can wright a representation as a formal series
\beq{2.18b}
H^{-1}=(\al^*)^{-1}\sum_{k=0}^{\infty}(-1)^kq^{-2k}((\al^*)^{-1}\ga^*)^k
(\ga\al^{-1})^k\al^{-1}.
\eq
It can be proved by the induction that
\beq{2.19}
H^l=(\al^*)^l\sum_{k=0}^lq^{(l-1)k}
\frac
{[l]_{q^2}!}
{[k]_{q^2}![l-k]_{q^2}!}
y^{*k}y^k\al^l,~l\geq 0,~~y=\ga\al^{-1},
\eq
$$H^{-l}=(\al^*)^{-l}\sum_{k=0}^{\infty}(-1)^kq^{-(l+1)k}
\frac
{[l+k-1]_{q^2}!}
{[k]_{q^2}![l-k]_{q^2}!}
y^{*k}y^k\al^l,~l> 0.$$
Here we use the standard notations
\beq{2.19a}
[k]_{q^2}=\frac{q^k-q^{-k}}{q-q^{-1}}=\frac{\sinh k \hbar_2}{\sinh \hbar_2},
\eq
where $q=\exp (-\hbar_2)
\footnote
{The subscribe 2
 is used here to distinguish this constant from the usual Planck constant
$\hbar_1$,
 which presents
 from the very beginning in front of derivatives in the classical group
approach.}$,
and $[k]_{q^2}!=[k]_{q^2}[k-1]_{q^2}\ldots 1$.

The normal order for  $z$ and $z^*$ can be derived  from
(\ref{2.18}) and (\ref{2.18b})
\beq{2.20}
z=\xi+\ka,~~\xi=\al^{-1}\be,
\eq
and
\beq{2.21}
\ka=\ga^*H^{-1}\al^{-1}
=\sum_{k=0}^{\infty}(-1)^kq^{-2k}(y^*)^{k+1}y^k\al^{-2},~~y=\ga\al^{-1}.
\eq
Also
$$\xi\ka^k=q^{2k}\ka^k\xi+(q^{2k}-1)\ka^{k+1},$$
and therefore
\beq{2.21a}
z^n=(\xi+\ka)^n=\sum_{l=0}^nq^{l(n-1)}
\frac
{[n]_{q^2}!}
{[l]_{q^2}![l-n]_{q^2}!}
\ka^l\xi^{n-l}.
\eq
{}From (\ref{2.20})
$$(z).A=q^{-1}z,~(z^*).A=z^*,$$
\beq{2.22}
(z).B=q^{-1/2},~(z^*).B=0,
\eq
$$(z).C=-q^{1/2}z^2,~(z^*).C=q^{1/2}H^{-2},$$
$$(z).D=qz,~(z^*).D=qz^*.$$
In the same way from (\ref{2.19}) for any $r\in{\bf Z}$
$$(H^r).A=q^{r/2}H^r,~(H^r).B=0,$$
\beq{2.23}
(H^r).C=q^{\frac{r-1}{2}}[r]_{q^2}H^rz,
\eq
$$(H^r).D=q^{-r/2}H^r.$$
 Formally, we can consider this relation for any $r\in{\bf C}$, since $H$
(\ref{2.19}) can be defined as a positive definit Hermitian operator.

The  relations
 (\ref{2.22}) and (\ref{2.23}) demonstrate the action of $U_q(SL_2(\bf {C}))$
on
 the horospherical coordinates of ${\bf L}_q^3$. Moreover, using the same
notations for this representation as for the generators of $U_q(SL_2(\bf{C}))$,
we can prove that in the classical limit (\ref{2.22}) and (\ref{2.23}) are
transformed in the Lie derivatives (\ref{1.4})
$$
d_A=\lim_{q\rightarrow 1}\partial_qA,~~d_D=\lim_{q\rightarrow 1}\partial_qD,$$

$$d_B=\lim_{q\rightarrow 1}B,~~d_C=\lim_{q\rightarrow 1}C.$$
As for the quantum Casimir (\ref{2.14}), it is easy to check that
\beq{2.24}
\lim_{q\rightarrow 1}\Omega_q=\frac{1}{2}\Omega+\frac{1}{4},
\eq
 ($\Omega$ is the classical Casimir (\ref{1.7})), as it has to be.

\section{Eigenfunctions of the quantum Casimir and scattering in spin chains}
The eigenvalue problem
\beq{3.1}
(f(H,z,z^*)).\Omega_q=-\lm^2f(H,z,z^*),
\eq
where $f(H,z,z^*)$ is a Laurent series in $H,z,z^*$,
can be solved by means of the relations from the previous section
(\ref{2.15a}) and (\ref{2.15b}).

In the similar fashion as in the classical case we choose a special form of
$f(H,z,z^*)$. Let
$$e_{q}^v=\sum_{n=0}^{\infty}\frac{v^n}{(q;q)_n},$$
where
\[(a;q)_n=\left\{\begin{array}{ll}
1&n=0,\\
(1-a)(1-aq)\ldots (1-aq^{n-1})&n>0,
\end{array}
\right. \]
be the q-exponent \cite{GR}. Then we assume
$$f(H,z,z^*)=e_{q^{2}}^{i\mu q^{-\frac{3}{2}}(1-q^2)z^*}F_{\lm}(H)
e_{q^{2}}^{i\mu q^{-\frac{3}{2}}(1-q^2)z},$$
and $F_{\lm}(H)=\sum_ra_{r+1}(\lm)H^r.$
Thus
\beq{3.2}
f(H,z,z^*)=\sum_{m=0}^{\infty}\sum_r\sum_{n=0}^{\infty}\frac
{(i\mu)^{m+n}q^{-m-n}(1-q^2)^{m+n}}
{(q^2,q^2)_m(q^2,q^2)_n}
a_{r+1}(\lm)X^{*m}H^rX^n,
\eq
where $X=q^{-\frac{1}{2}}z$.

{}From (\ref{2.21a})
\beq{3.3}
v^{(m,r,n)}=X^{*m}H^rX^n=
\eq
$$q^{\frac{m+n}{2}}\sum_{k=0}^m\sum_{l=0}^n
\frac
{q^{k(m-1)+l(n-1)}
[m]_{q^2}![n]_{q^2}!}
{[k]_{q^2}![m-k]_{q^2}![l]_{q^2}![n-l]_{q^2}!}
\xi^{*m-k}\ka^{*k}H^r\ka^l\xi^{n-l}.$$
As it follows from (\ref{2.1})
$$\ka^{*k}=(\al\ga)^{*-k}\ka^k(\al\ga)^k.$$
It can be find from (\ref{2.15b}) and (\ref{2.16}) for $n\geq 0$
$$(\xi^n).A=q^{-n}\xi^n,$$
$$(\xi^n).B=q^{-1/2}[n]^{q^2}\xi^{n-1},$$
$$(\xi^n).C=-q^{1/2}[n]_{q^2}\xi^{n+1},$$
$$(\xi^n).D=q^{n}\xi^n,$$
For $n\geq 0$ after some algebra we obtaine
$$(\ka^n).A=q^{-n}\ka^n,$$
$$(\ka^n).B=0,$$
$$(\ka^n).C=-q^{n+1/2}[2n]_{q^2}\ka^n\xi-q^{2n+1/2}[n]_{q^2}\ka^{n+1},$$
$$(\ka^n).D=q^n\ka^n.$$
Then the following relation
$$(\al\ga)^{-1}=\ka^{*-1}H^{-2}+\ka$$
allows to derive the actions of the generators on the monom
$w^{(k,r,l)}=\ka^{*k}H^r\ka^l$
$$(w^{(k,r,l)}).A=q^{r/2-l}w^{(k,r,l)},$$
$$(w^{(k,r,l)}).B=0,$$
$$(w^{(k,r,l)}).C=q^{-k+l+\frac{r+1}{2}}[k]_{q^2}w^{(k-1,r-2,l)}-$$
$$-q^{l-\frac{r-1}{2}}
[2l-r]_{q^2}w^{(k,r,l)}\xi-q^{2l-\frac{r-1}{2}}[l-r]_{q^2}w^{(k,r,l)+1},$$
$$(w^{(k,r,l)}).D=q^{-r/2+l}w^{(k,r,l)}.$$
In agreement with (\ref{3.3})
$$(v^{(m,r,n)}).A=q^{\frac{r}{2}-n}v^{(m,r,n)},$$
$$(v^{(m,r,n)}).B=q^{r/2}[n]_{q^2}v^{(m,r,n-1)},$$
$$(v^{(m,r,n)}).C=q^{m+n=\frac{r-2}{2}}[m]_{q^2}v^{(m-1,r-2,n)}-$$
$$-q^{-r/2}[n-r]_{q^2}v^{(m,r,n+1)},$$
$$(v^{(m,r,n)}).D=q^{-\frac{r}{2}+n}v^{(m,r,n)}.$$
Using (\ref{2.15a}) and (\ref{2.15b}), we obtaine
$$(v^{(m,r,n)}).\Omega_q=[\frac{r+1}{2}]^2_{q^2}v^{(m,r,n)}
+q^{m+n+r-2}[m]_{q^2}[n]_{q^2}v^{(m-1,r-2,n-1)}.$$
Substituting it in (\ref{3.2}) we come to
$$(f(H,z,z^*)).\Omega_q=e_{q^{2}}^{i\mu q^{-\frac{3}{2}}(1-q^2)z^*}\sum
_ra_{r+1}[\frac{r+1}{2}]^2_{q^2}H^re_{q^{2}}^{i\mu q^{-\frac{3}{2}}(1-q^2)z}-$$
$$-\mu^2e_{q^{2}}^{i\mu q^{-\frac{3}{2}}(1-q^2)z^*}\sum
_ra_{r+1}q^{r-2}H^{r-2}e_{q^{2}}^{i\mu q^{-\frac{3}{2}}(1-q^2)z}.$$
Therefore, (\ref{3.1}) is reduced to
$$\sum_ra_{r+1}[\frac{r+1}{2}]^2_{q^2}H^r-\mu^2\sum_ra_{r+1}q^{r-2}H^{r-2}=$$
$$-\lm^2\sum_ra_rH^r.$$

This relation is equivalent to our main eqution
\beq{3.4}
\frac
{F_{\lm}(qH)-2F_{\lm}(H)+F_{\lm}(q^{-1})}
{(q-q^{-1})^2}-\mu^2q^{-3}
H^{-2}F_{\lm}(qH)=-\lm^2F_{\lm}(H).
\eq
It is the lattice counterpart of the Liouville equation (\ref{1.10}) -
in the limit $q\rightarrow 1,
{}~(\hbar_2\rightarrow 0)$ it is transformed in (\ref{1.10}).
 Here
the second derivative is replaced on the second difference operator. This
naive modification
of (\ref{1.10})
 can be easily predicted from the very beginning.
 The less obvious modification is the shift
 of the argument in the potential operator. It is of course
will be crucial for the solving the eigenvalue problem exactly.

Using the shift operator in the form
$$\exp (\hbar_2\partial) F(H)=F(qH),$$
we can rewrite the last relation as
\beq{3.4a}
[\frac{(\sinh )^2(\frac{ \hbar_2\partial}{2})}{(\sinh )^2\hbar_2}-\mu^2q^{-3}
H^{-2}e^{\hbar_2\partial}] F_{\lm}(H)=-\lm^2F_{\lm}(H).
\eq

To solve (\ref{3.4}) assume that
\beq{3.5}
\mu^2q^{-5}(q-q^{-1})=1,
\eq
and replace the eigenvalue parameter
\beq{3.6}
\lm=[i\th]_{q^2}=\frac{\sin (\frac{ \hbar_2\th}{2})}{\sinh \hbar_2},
{}~~x=\cos \hbar_2\th,
\eq
Since (\ref{3.4}) is the difference operator, put
\beq{3.6a}
H=q^{-n},
\eq
\beq{3.7}
c_n(x)=\frac{1}{(q^2,q^2)_n}F_{\lm}(H)|_{H=q^{-n}}.
\eq
In the new variables (\ref{3.4}) takes the form
\beq{3.7b}
(1-q^{2n+2})c_{n+1}(x)+c_{n-1}(x)=2xc_n(x).
\eq
This equation has the form of reccurence relation for orthoganal polynomials.
If we put $c_{-1}=0$ and $c_0=1$,
 then (\ref{3.7}) defines the q-continuos Hermit
polynomials, which are particular case of the Rogers-Askey-Ismail polynomials
\cite{GR}, or, as the same,
 the Macdonald polynomials for the $A_1$ root system \cite{M}.
Namely,
\beq{3.7a}
c_n(\cos \hbar_2\th)=C_n(\cos \hbar_2\th;t|q^2)|_{t=0},
\eq
where $C_n(\cos \hbar_2\th;t|q^2)$ is Rogers-Askey-Ismail polynomial
$$C_n(\cos \hbar_2\th;t|q^2)=\sum_{a+b=n, a,b\in \bf{Z}^+}
\frac
{(t;q^2)_a(t;q^2)_b}
{(q^2;q^2)_a(q^2;q^2)_b}
e^{i\hbar_2\th (a-b)},$$
and for the Macdonald polynomials $P_n$
\beq{3.8}
F_{\lm}(H)|_{H=q^{-n}}=(q^2;q^2)_n c_n(\cos \hbar_2\th)=
\eq
$$(t;q^2)_nP_n(e^{i\hbar_2\th} ;t|q^2)|_{t=0}.$$

 It is
worthwhile to note, that the degrees of polynomials $n$ play now the role
 of the "space" variable (\ref{3.7}), while their arguments are related
 to the eigenvalue parameter $\lm$
 (\ref{3.6}), (\ref{3.6a}) .
 Note that the argument $\th$, which defined the energy
$E=\lm^2/2$ by (\ref{3.6})
 lies in the interval $\th\in[0,\frac{4\pi}{\hbar_2})$.
 The similar phenomena arises also for
the zonal spherical functions
 on quantum "noncompact" symmetric spaces (see \cite{Z,FZ},
where the case of $U_q(SU(1,1))$ was considered).

The asymptotic behavior
 of $c_n$ for $n\rightarrow \infty$ is following \cite{GR}
\beq{3.9}
(q^2;q^2)_n c_n(\cos \hbar_2\th)\sim\{\frac
{q^{-in\th}}{(q^{2i\th};q^2)_{\infty}}+
\frac
{q^{in\th}}{(q^{-2i\th};q^2)_{\infty}}
\}.
\eq
Thus, the Harish-Chandra $c$-function is equal to
$$c(\lm)=\frac{1}{(q^{-2i\th};q^2)_{\infty}}.$$

To find a spin chain,  which exhibits the same scattering, it is convenient to
consider the extended theory and switch on the additional parameter $t$
(\ref{3.7a}), (\ref{3.8}).
This theory was considered in \cite{FZ}. It was demonstrated there that the
Harish-Chandra $c$-functions for the generic $A_1$ Macdonald polynomials
gives rise for the scalar $S$-matrix to the scattering of two special
excitations in $Z_N$-Baxter model \cite{B,G} in the $N\rightarrow\infty$
regime.
It has two parameters -
 the modular parameter $\tau$ and the anisotropy parameter
$\ga$.
In fact, $\tau$ is a modular parameter of an elliptic curve and $\ga$ is a
point
on it. They are parameters of the Sklyanin algebra \cite{S,Ch,OF}, which allows
a solution of corresponding Yang-Baxter equation.
The parameters of the Macdonald polynomials are related to later as
follows
\beq{3.11}
q=e^{2i\pi\tau},~t=q^{\frac{i\ga}{\pi\tau}}.
\eq

Some interesting solutions corresponding to
points in the space $(\tau,\ga)$ were considered in \cite{FZ}.
In our case $t=0$.
  Since $\Im m\tau >0$ and $\Re e\tau=0$,  then $\ga\rightarrow\infty$.
 Simultaneously $N\rightarrow\infty$.

 As it follows from
the relation (5.2) \cite{FZ},
 there is the duality in the scattering picture in this model
$$(N,\frac{-i\pi\tau}{\ga})\sim(\frac{-i\pi\tau}{\ga},N).$$
Thus the same $S$-matrix can be realized in the $Z_N$-Baxter model in the
regime
$$\ga\rightarrow 0,~N\rightarrow 0.$$

\section{Conclusion}
We have found that the wave functions of the q-Liouville Hamiltonian lie in
the same family of Macdonald polynomials as the wave functions of the
q-Sutherland Hamiltonian \cite{Z,FZ,Ko}.
In the former case $t=0$ and $t=q$ in the later (see (\ref{3.8})).
 The Macdonald polynomials have an interpretation as
 the "zonal spherical functions" on the Sklyanin quantum algebra \cite{FZ}.
Thus there exists the transition from
the usual $q=1$ Sutherland Hamiltonian
 to the Liouville quantum mechanics through
the Sklyanin quantum algebra.
 It is natural to conjecture that this transition can be generalized
 on the arbitrary number of particles.
Recall that the both Hamiltonians can be also considered as two special
 reductions of Casimirs on $U_q(SL_2),~q\leq 1$. But this unification is
performed in the different way by the increasing  of  degrees of freedom.

  One of the way to study quantum integrable systems is to put
 the space variable on a lattice. In particular, for the Liouville field
theory it was done firstly in \cite{FT}, at least quasiclassically. Our
approach
is differ from it - instead of the discretization of the space variable we
 discretized the zero modes of the Liouville field.
 It is similar to the substitution
a continuos gauge group on a discrete subgroup in the Monte-Carlo simulation
of gauge theories.
Introducing the lattice is
equivalent to introducing an infinite hard wall instead
 of the Liouville exponential
potential - the wave functions are vanish for $n\leq 0$.
The theory, as a modified version of the zero modes dynamics of 2d gravity,
is still ultraviolet free since for small distances  the potential
vanishes $H^{-2}=q^{2n}_{n\rightarrow +\infty}=0$
 (see (\ref{3.4}),(\ref{3.6a}).

  On the other hand the system under consideration resembles the Ruijsenaars
relativistic Toda model \cite{R} for two particles,
 though it does not coincide with it .
In spite of later it allows to solve it exactly . It
 is possible also to find the
explicit solution in the classical case after the substitution $i\partial
\rightarrow p$. Unfortunately, this solution has not clear group-theoretical
interpretation, as the  solutions of
 the open nonrelativistic Toda model \cite{OP}.
At the same time the Ruijsenaars solutions are natural $q$-deformation of
the non relativistic Toda solutions.  It will be interesting to find
a similar interpretation for the solution coming from the quantum Lorentz
group.

Starting from the Hamiltonian (\ref{3.4}) it is easy to guess the form of the
lattice N-body Toda quantum mechanics. But it will not be easy to justify this
Hamiltonian by means of the Iwasawa decomposition of ${\cal A}_q(SL(N,{\bf
C}))$
\cite{JS} using the same type calculations as above. Nevertheless,
investigations of $q$-Whitteker functions in a general case and their
 relations with representations of quantum groups is plausible.

The next desirable generalization of this model is its two-dimensional version
in the quantum and classical form. Note that one of the possible versions
of the q-deformed classical Liouville field theory  can be reconstructed
in principle as in the classical theory from
the q-deformed WZW model, proposed, for example, in \cite{LS}  .

\bf Acknowledgments. {\sl We are grateful to A.Zabrodin for  numerous valuable
  comments. Part of this work was done in the Aspen Center for Physics and
in the INFN sezione di Ferrara. M.O. would like to thank Andre Leclair
,  Dennis Nemeshansky
 and Gianni Fiorentini for their hospitality. Discussions
 with N.Reshetehin , F.
Smirnov and other participants of the Aspen Workshop were very useful.}
\small{

}

\end{document}